\documentclass[runningheads]{llncs}
\usepackage{graphicx}
\usepackage{enumitem}
\usepackage{url}
\usepackage[linesnumbered,ruled,vlined]{algorithm2e}
\usepackage{caption}
\usepackage{orcidlink}
\usepackage{subcaption}

\begin{document}
\title{AI Potentiality and Awareness: A Position Paper from the Perspective of Human-AI Teaming in Cybersecurity}
%
\titlerunning{AI Potentiality and Awareness}
%
%
\authorrunning{Sarker et al.}
%
%

\author{Iqbal H. Sarker\textsuperscript{1,2,*}\orcidlink{0000-0003-1740-5517}, Helge Janicke\textsuperscript{1,2}\orcidlink{0000-0002-1345-2829}, Nazeeruddin Mohammad\textsuperscript{3}\orcidlink{0000-0003-3580-5960}, Paul Watters\textsuperscript{4}\orcidlink{0000-0002-1399-7175}, Surya Nepal\textsuperscript{2,5}\orcidlink{0000-0002-3289-6599}}
%
\authorrunning{Sarker et al.}
%
\institute{\textsuperscript{1} Security Research Institute, Edith Cowan University, Perth, WA 6027, Australia.\\ \textsuperscript{2} Cyber Security Cooperative Research Centre, Australia. \\
\textsuperscript{3} Cybersecurity Center, Prince Mohammad Bin Fahd University, Khobar, SA. \\
\textsuperscript{4} Cyberstronomy Pty Ltd, Melbourne, VIC, Australia. \\\textsuperscript{5}
Data61, CSIRO, Sydney, NSW 2122, Australia.\\
$^*$Correspondence: m.sarker@ecu.edu.au
}
\maketitle              
\begin{abstract}
This position paper explores the broad landscape of AI potentiality in the context of cybersecurity, with a particular emphasis on its possible risk factors with awareness, which can be managed by incorporating human experts in the loop, i.e., ``Human-AI" teaming. As artificial intelligence (AI) technologies advance, they will provide unparalleled opportunities for attack identification, incident response, and recovery. However, the successful deployment of AI into cybersecurity measures necessitates an in-depth understanding of its capabilities, challenges, and ethical and legal implications to handle associated risk factors in real-world application areas. Towards this, we emphasize the importance of a balanced approach that incorporates AI's computational power with human expertise. AI systems may proactively discover vulnerabilities and detect anomalies through pattern recognition, and predictive modeling, significantly enhancing speed and accuracy. Human experts can explain AI-generated decisions to stakeholders, regulators, and end-users in critical situations, ensuring responsibility and accountability, which helps establish trust in AI-driven security solutions. Therefore, in this position paper, we argue that human-AI teaming is worthwhile in cybersecurity, in which human expertise such as intuition, critical thinking, or contextual understanding is combined with AI's computational power to improve overall cyber defenses.

\end{abstract}
\section{Introduction}
In today's rapidly evolving digital technology ecosystem, cybersecurity has taken on unprecedented significance. With the growth of interconnected systems, protecting sensitive information and critical infrastructure has become a top priority of a nation \cite{sarker2022machine}. As threats to digital assets grow in complexity and sophistication, there is an urgent need for innovative approaches to effectively counter these concerns. In this position paper, we focus on the potential of artificial intelligence (AI) as well as its awareness, i.e., to handle possible risk factors, strengthened by human expertise and observation in real-world application areas. 

The combination of AI with human expertise has emerged as a potential solution to addressing the ever-changing landscape of cybersecurity. AI systems have shown promise in automating repetitive tasks, analyzing large datasets, and discovering patterns that may be beyond human observation \cite{zhang2022artificial} \cite{sarker2022multi}. These competencies are particularly beneficial in the field of cybersecurity, where the rapid identification and mitigation of threats and attacks are crucial. Human experts, on the other hand, have a distinct set of abilities in their understanding of context, intuition, and ethical judgment. This position paper intends to dive into the intricate dynamics of this Human-AI collaboration, exploring the synergies between AI's computational capability and human experts' domain knowledge, contextual reasoning, and critical thinking in their roles. 

The cooperation between AI and human expertise in cybersecurity is not only an issue of technological collaboration; it also extends to the ethical sphere. The successful deployment of AI in cybersecurity operations necessitates resolving issues of bias, accountability, transparency, and the right to distinguish decision-making authority between machines and humans. Human analysts can make sophisticated decisions depending on the broader context of the organization's goals and risk tolerance. Thus, human observations at the application level are still required despite AI's huge potential in computing \cite{sarker2023data}. In short, we define ``AI potentiality" as \textit{technical aspects} of AI, emphasizing its computational capabilities in terms of speed, accuracy, scalability, and automation in cybersecurity tasks, and ``AI awareness" as its \textit{possible risk factors} highlighting the importance of incorporating human experts in the loop, e.g., human domain knowledge, intuition, and situational awareness to make informed decisions in cybersecurity, which eventually help to achieve organizations goal with proper responsibility and accountability. Thus ``Human-AI teaming", in essence, maximizes the assets of both AI and human intelligence, resulting in a more robust, effective, and explainable security system for organizations. To better comprehend the core topic of this position paper and overall contributions, we formulate three key questions below:

\begin{itemize}
     \item \textit{AI Potentiality:} Is AI capable of effectively and efficiently processing, analyzing, and extracting insights or useful knowledge from massive amounts of cyber data, which is challenging for a human analyst?
     
    \item \textit{AI Awareness:} Are risk factors associated with real-world AI applications in the context of cybersecurity, particularly how we can ensure responsibility and accountability if the AI system fails in certain situations?

    \item \textit{Human-AI Teaming:} Is it worthwhile to establish human-AI teaming for cybersecurity solutions and rethink the current cyberspace making informed decisions with accountability in real-world application areas?
\end{itemize}

Overall, this position paper emphasizes a deeper understanding of Human-AI teaming, particularly how the combination of AI and human expertise might strengthen our cyber defense measures in various real-world application areas. In the following sections, we provide an explicit understanding of the potential of AI, as well as the importance of human experts in the loop in the context of next-generation cybersecurity solutions. Therefore, this paper aims to provide significant insights for cybersecurity academics, practitioners, policymakers, and stakeholders, advancing discussions that lead the way for successful and responsible cybersecurity practices in our digitally connected society.

\section{Why Human-AI Teaming in Cybersecurity?}
\label{Why AI-Powered Solutions in Cybersecurity?}
In cybersecurity, the term ``Human-AI teaming" refers to cooperation between AI technologies and human expertise to boost the overall effectiveness of cybersecurity solutions. In the following, we discuss how this teaming can contribute:

\begin{itemize}
    \item \textit{Complementary Strengths:} AI is particularly effective at swiftly analyzing and processing large volumes of data, finding patterns, discovering anomalies, and generating policies from complex specifications. However, there should be clear accountability for decisions made by AI systems in real-world cybersecurity applications. Human experts, on the other hand, have awareness of the environment, intuition, and the skills to make complex decisions using a combination of technological knowledge and hands-on expertise. Thus, organizations can achieve better cyber attack or threat analysis and response strategies with proper accountability through this integration.

    \item \textit{Scale and Speed:} 
    Cyber threats are constantly evolving and can emerge at a rapid pace. Attackers often employ automated tools and methods to exploit vulnerabilities. AI can assist cybersecurity teams in analyzing threats and taking action at a speed and scale that would be impossible for humans to tackle by themselves. This speeding up is essential for minimizing potential harm and promptly addressing emerging risk factors. 

    \item \textit{Decision Support:} Advanced threats often consist of numerous stages and sophisticated attack pathways. AI can give human analysts pertinent data and insights to support their decision-making. This can include information about the threat's nature, its potential effects, and recommended solutions. While humans can handle the more complicated and hidden aspects of investigations, attribution, and response, AI can discover insights from data as well as automate the necessary modules of threat detection and analysis. 

    \item \textit{User Behavior and Predictive Analytics:} AI can monitor user behavior patterns to find anomalies that could point to unauthorized access or compromised accounts. Analyzing historical data AI can identify trends and predict potential future threats, which eventually enables human analysts to take proactive measures before an incident happens.

    \item \textit{Continuous Learning and Improvement:} 
    To effectively respond to new threats human analysts can regularly update and improve AI models. Although AI models are capable of learning from previous data, they may require human supervision to understand the importance of newly discovered patterns and anomalies. This teaming enables a feedback loop where human experts can instruct and enhance AI models based on their real-world experiences, resulting in progressively more precise and efficient threat analysis.

    \item \textit{Incident Response and Recovery:} When a cyber incident occurs, human expertise is essential for managing the situation, determining the level of the damage, coordinating the response, and communicating with stakeholders. AI can assist in quickly analyzing incident data, offering pertinent insights to guide human decision-making as well as necessary automation. This helps human experts for better strategic decisions, determine how an incident may affect other stakeholder groups, and efficiently direct the recovery process.

    \item \textit{Regulatory Compliance, Trust and Accountability:} In many industries, security operations are subject to regulations that require human oversight and accountability. Particularly, need to ensure who will be responsible if AI systems fail. Stakeholders often want to understand the rationale behind security decisions. Cybersecurity experts can provide explanations for AI outcomes that build trust, transparency, and accountability.
\end{itemize}

Although AI has huge potential, it may have risk factors, particularly in terms of situational awareness, decision-making, and accountability, as highlighted above. Thus, to increase overall cybersecurity resilience, it needs to blend human intuition, critical thinking, and decision-making skills with AI's speed, scalability, data processing and learning capabilities.

\section{AI-based Modeling and Potentiality}
\label{AI-based Modeling}
In this section, we first highlight what types of AI-based modeling can be built, and then we explore multi-aspects AI methods that can contribute in the context of cybersecurity modeling.

\subsection{Major Types of AI Modeling}
Generative AI and discriminative AI are two common approaches in the area, with hybrid AI being a combination of the two. Generative AI is typically focused on generating new data according to needs, while discriminative AI is focused on classifying data. Hybrid AI combines these two approaches to take advantage of their strengths to solve a particular problem. In the following, we explore these with examples in the context of cybersecurity.

\begin{figure}
\centering
\begin{minipage}{.5\textwidth}
  \centering
  \includegraphics[width=.95\linewidth, height=3cm]{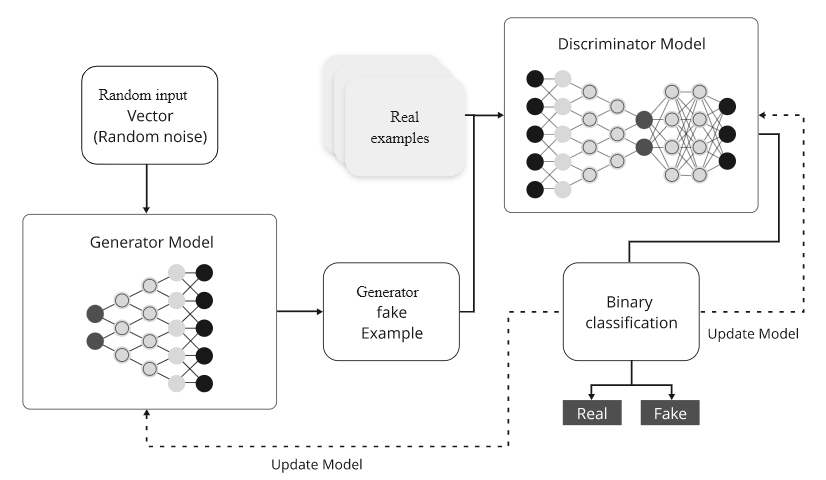}
  \captionof{figure}{Generative Adversarial Network}
  \label{label:Generative-AI}
\end{minipage}%
\begin{minipage}{.5\textwidth}
  \centering
\includegraphics[width=.95\linewidth, height=3cm]{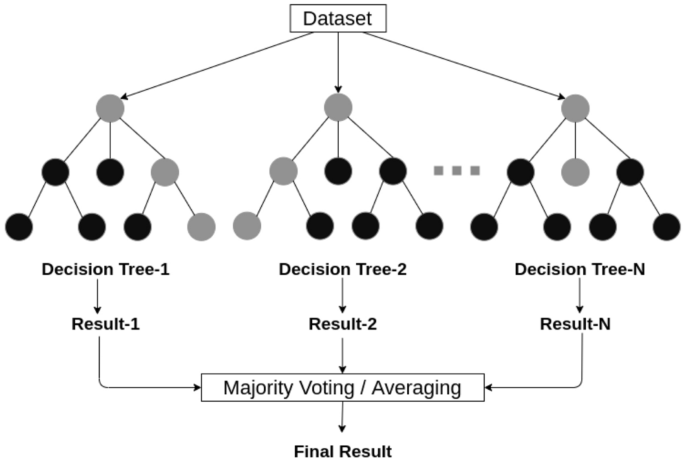}
  \captionof{figure}{Random Forest}
  \label{label:Random forest}
\end{minipage}
\end{figure}

\begin{itemize}
    \item \textit{Generative AI:} Generative models are trained to learn the underlying structure of the data and generate new data that has similar characteristics to the original data, e.g., Generative Adversarial Network (GAN) \cite{yinka2020review}, as shown in Fig \ref{label:Generative-AI}. In cybersecurity, generative models can be used for various tasks such as attack simulation to test and enhance defense mechanisms,  synthetic data generation for training machine learning models, anomaly detection, etc.

    \item \textit{Discriminative AI:} Discriminative AI models aim to learn the decision boundary between different classes of data, e.g., Random Forest (RF) with multiple decision trees \cite{kilincer2021machine}, as shown in Fig \ref{label:Random forest}. In cybersecurity, discriminative models can be used for various tasks such as intrusion detection, predicting threats, fraud detection, malware analysis, etc.

    \item \textit{Hybrid AI:} Hybrid models combine the strengths of both generative and discriminative models. For example, a hybrid model could use a generative model to generate synthetic data that can be used to train a discriminative model for a particular task \cite{sarker2022multi}. In cybersecurity, hybrid AI can be used to create more robust intrusion detection systems, detect advanced persistent threats, simulate complex attack scenarios, etc.
\end{itemize}

Overall, each type of AI modeling mentioned has its own strengths and weaknesses, and the best approach for a given task depends on the specific use case and availability of the cyber data and relevant resources. A combination of generative, discriminative, and hybrid AI approaches can be effective in providing a comprehensive and effective cybersecurity solution by taking into account their individual strengths. Thus, research should be focused in this area for future-generation cybersecurity modeling.

\subsection{AI Methods and Algorithms}
Although AI is a broader term in terms of techniques and application areas, we explore the key AI methods that can be used in cybersecurity according to the problem nature and available data or resources. These are:

\begin{itemize}
    \item \textit{Machine Learning: } Machine learning has emerged as a powerful tool in cybersecurity. Machine learning algorithms such as Decision Trees, RF, SVM, Logistic regression, Clustering, PCA, etc. \cite{sarker2022machine} can analyze large volumes of data, such as network traffic and system logs, to identify patterns and anomalies that could indicate a potential cyber attack. By analyzing attack patterns, machine learning algorithms can identify the root cause of security breaches, enabling security teams to take corrective action more quickly and effectively.

    \item \textit{Deep Learning:} Deep learning is a subset of machine learning that involves training neural networks with multiple layers to learn from complex data sets \cite{sarker2022multi}. Deep learning models such as DNN, CNN, RNN, LSTM, Autoencoder, GAN, etc. \cite{sarker2022machine} have the ability to extract representational information and decision rules simultaneously from data \cite{dick2019deep}. Deep learning algorithms can be used in cybersecurity, particularly in the areas of threat detection, anomaly detection, malware classification, and so on \cite{sarker2022machine} \cite{dixit2021deep}.

    \item \textit{Semantic Knowledge Representation and Reasoning:} Semantic knowledge involves a deeper understanding of the evolving threat landscape and the context of data, which eventually helps security professionals make informed decisions in real time. Several popular methods such as ontologies or knowledge graphs, i.e., graph-structured data model, can be used to represent knowledge about potential threats, vulnerabilities, and assets in cybersecurity \cite{valja2020automating}. 

    \item \textit{Knowledge or Rule Discovery:} Knowledge discovery, e.g., rules, in cybersecurity typically involves extracting useful insights and patterns from large datasets that are used in intrusion detection, access control, and other security systems. Rule-based AI models can leverage machine learning and data science processes to enhance security solutions by generating a set of optimized rules. For instance, clustering algorithms can be used to group similar behavioral patterns of network traffic and create rules accordingly to detect anomalies in real time. Association rule mining techniques can assist security analysts in uncovering relationships between various events and entities involved in the incident \cite{sarker2022multi}. 

    \item \textit{Language Model and Multimodality:} Large language models (LLM) are typically at the forefront of natural language processing (NLP) research and applications due to their effectiveness in understanding and generating human language. These models can analyze and understand vast amounts of unstructured text data, such as threat reports to extract relevant threat intelligence. For example, to summarize long threat reports and classify them into different threat categories these models can be used, which eventually assist security analysts in prioritizing their responses. In addition, multimodal intelligence by taking into account the fusion of textual, visual, or sensor data \cite{bakalos2019protecting} can lead to more robust and comprehensive solutions in cybersecurity depending on relevant data availability. 
\end{itemize}

Overall, these multi-aspects AI methods and algorithms discussed as well as their hybridization can provide powerful tools for detecting and responding to security threats. However, it is important to carefully consider the strengths and weaknesses of each method and to use them appropriately based on the specific needs and characteristics of the systems being protected.

\section{Real-World Cybersecurity Application Areas}
In the real-world application areas, both IT (Information Technology) and Industrial Control Systems (ICS) or Operational Technology (OT) systems are important to the operation of modern organizations and business \cite{sarker2023data}. Human-AI teaming is crucial for ensuring the security of IT and OT systems, networks, and data from cyber threats. Hence, we summarize the potential usage of human-AI teaming, which can be applicable for both IT and ICS/OT security systems:

\begin{itemize}
    \item \textit{Anomaly and Threat Detection with Real-time Monitoring:} AI can continuously monitor network traffic, system logs or sensor data to identify anomalies or suspicious activity that might indicate a cyberattack. Human experts can then evaluate and investigate these findings and take necessary action.
    
    \item \textit{Incident Response and Recovery}:
     AI can classify and prioritize security alerts depending on severity, enabling human responders to focus on the most critical incidents. AI may automate predefined incident response playbooks, reducing risks immediately while notifying human teams for further investigation, and strategic decision-making.

    \item \textit{Risk Assessment and Vulnerability Management:} AI can analyze vulnerabilities and risks in both IT and ICS/OT environments by continuously investigating configurations, system states, and emerging threats. Security analysts can assess and prioritize risks and vulnerabilities for remediation depending on the possible impact on critical infrastructure.
    
    \item \textit{Predictive Analytics and Proactive Maintenance:} AI can analyze historical data and current trends to anticipate potential future threats or vulnerabilities. Human experts use this predictive insight to resolve vulnerabilities and security weaknesses in advance. In many cases, these predictive AI insights can assist in designing maintenance plans and making informed decisions about equipment replacements or upgrades.

    \item \textit{Incident Investigation and Forensics:} AI assists in capturing and analyzing forensic data to determine the underlying causes of security problems. Human investigators employ AI-generated insights to build together attack narratives, identify criminals, and attribute breaches.

    \item \textit{Adaptive Security:} AI-powered adaptive security mechanisms adjust dynamically based on real-time threat intelligence and system conditions. Human analysts supervise and fine-tune AI algorithms to ensure they align with business goals and regulatory constraints.

    \item \textit{Visualization and Situational Awareness:} AI analyses and visualizes data, allowing human operators to instantly understand the security status. In critical scenarios, human analysts investigate AI-generated visualizations to make well-informed decisions.

    \item \textit{Threat Intelligence and Contextualization:} AI is capable of analyzing massive amounts of threat data and providing actionable intelligence. Human analysts contextualize threat intelligence, making it applicable to specific contexts.

    \item \textit{Threat Hunting:} Human analysts can employ AI-powered technologies to proactively hunt for evidence of compromise that may not trigger automatic alarms. Human analysts' intuition and expertise are essential in uncovering sophisticated attacks. Overall, humans can develop the hypothesis and AI can help to execute it.

    \item \textit{Behavioral Analysis and User Authentication:} AI can develop behavioral profiles for individuals and groups inside an organization. Any variations from these characteristics can generate alerts to explore potential insider threats. Human experts can evaluate suspicious behaviors to identify whether they are insider threats or false positives. Thus this can assist human experts in strengthening authentication systems of an organization.
\end{itemize}

Overall, Human-AI teaming in both IT and ICS/OT environment has the potential to greatly improve attack identification, incident response, and overall cybersecurity posture. AI can aid in automation, quick analysis, and processing of massive volumes of data, while human knowledge adds context, critical thinking, decision-making, and flexibility to a dynamic and evolving threat environment. However, the synergistic abilities of humans and AI need to be successfully exploited to handle domain-specific issues.

\section{Challenges and Research Direction}
\label{Challenges and Research Direction}
While AI-based cybersecurity solutions have the potential to enhance security as well as assist human experts in making decisions, several challenges need to be addressed to fully realize these benefits. Hence, we outline the key challenges and directions within the scope of our study:

\begin{itemize}
 \item \textit{Data Quality and Diversity:}  Getting tagged data with high quality and diversity to cover a variety of attack scenarios is a challenge. Thus research should focus on investigating techniques that can identify anomalies and potential risks by using both labelled and unlabeled data as well as unsupervised learning techniques, rather than relying solely on labeled data. In addition, methods to augment and synthesize a variety of representative datasets can help to achieve generalizability and robustness of AI models. 

 \item \textit{Adversarial Attacks:} Adversarial attacks against AI models in cybersecurity can lead to the evasion of detection systems by attackers. It is essential to conduct research to develop AI models that can resist adversarial attacks and maintain their effectiveness. Developing techniques such as adversarial learning, input sanitization, defensive distillation, and ensemble modeling could enhance model security.

 \item \textit{Interpretable AI:} To understand why certain decisions are made, it is crucial for cybersecurity analysts to comprehend how AI models make decisions. Thus research should focus on creating interpretable AI and machine learning models enabling cybersecurity professionals to validate and trust the model's outputs. The interaction between AI systems and human cybersecurity experts can be facilitated by developing innovative methods for explaining how AI-based cyber models make decisions.

 \item \textit{Innovative and Adaptive Modeling:}
 Traditional approaches have difficulty identifying zero-day vulnerabilities. A crucial research direction is the creation of AI models that can identify and mitigate previously unknown vulnerabilities. Examples of these models include anomaly detection, behavior analysis, and vulnerability prediction. Research is required to enhance models' ability to generalize novel and previously unknown attack patterns in different cybersecurity areas. Continuous learning is a key research direction for creating AI models that can adapt to dynamic attack tactics and changing threat environments. Effective threat detection can be improved exploring several modalities and comprehend temporal patterns, such as network traffic, system logs, and user behavior.

 \item \textit{Generative AI Modeling:}
 Generative AI has the potential to play a significant role in cybersecurity solutions by enhancing various aspects of threat analysis and response. For instance, generative AI can help in creating adversarial scenarios to evaluate the reliability of cybersecurity systems, assisting in the identification of vulnerabilities and the development of more robust defense mechanisms against such attacks. More research is needed to investigate techniques that enable AI models to generalize to unseen threats and adapt to evolving attack techniques.

 \item \textit{Privacy Concerns:} 
 Handling private and sensitive data could be a part of using AI in cybersecurity. It is challenging to achieve a balance between the advantages of AI and privacy issues while maintaining compliance with data protection laws. To resolve privacy issues, federated learning enables models to be trained across distributed devices without exchanging raw data. Investigating how distributed learning can be used effectively in cybersecurity to enable collaborative model training across various organizations, could be a potential area of research. 
 
 \item \textit{Regulatory and Ethical Frameworks:} To ensure secure and responsible deployment of AI-based model into cybersecurity, establishing guidelines, regulations, and ethical considerations are important. Investigating how AI may offer insights and recommendations while allowing people to make informed decisions is necessary. Research is needed to effectively integrate AI models into the decision-making procedures of cybersecurity analysts, which can enable synergistic interaction between human expertise and AI capabilities. 
 \end{itemize}

Although research has been significantly progressing in the area of AI, the challenges for effective security modeling still remain unaddressed. To advance the field of AI-based cybersecurity and enhance the general security posture of today's digital systems and interconnected networks, these identified issues and potential directions might help for next-generation cybersecurity solutions.

\section{Conclusion}
\label{Conclusion}
This position paper emphasized the importance of a balanced strategy that capitalizes on the strengths of both AI and human expertise, creating collaboration and trust between these two entities in the context of cybersecurity. As stated in this position paper, the synergy between human expertise and AI capabilities holds enormous promise for addressing the ever-changing landscape of cyber threats. We explored how AI technology might improve human capabilities by automating regular operations, analyzing and detecting anomalies at scale, as well as providing actionable insights for speedy decision-making. Furthermore, in terms of context understanding, intuition, accountability, and creativity in designing unique security solutions, the human element remains crucial. Overall, we believe that human-AI teaming can significantly improve the way we protect against cyber-attacks by providing a more resilient, efficient, and adaptive cybersecurity ecosystem. 

\section*{Acknowledgement}
The work has been supported by the Cyber Security Research Centre Limited whose activities are partially funded by the Australian Government’s Cooperative Research Centres Program.

\bibliographystyle{unsrt}
\bibliography{myref.bib}
\end{document}